\documentclass[twocolumn,showpacs,preprintnumbers,amsmath,amssymb]{revtex4}

\usepackage{graphicx}
\usepackage{dcolumn}
\usepackage{bm}

\begin{document}

\title{Scanning Tunneling Spectroscopy of the superconducting proximity effect in a diluted ferromagnetic alloy}

\author{L. Cr\'{e}tinon, A. K. Gupta\footnote{Present address : Department of Physics, Indian Institute of Technology, Kanpur 208016, India.}}
\affiliation{Centre de Recherches sur les Tr\`es Basses Temp\'eratures - C.N.R.S. and Universit\'e Joseph Fourier, 25 Avenue des Martyrs, 38042 Grenoble, France.}

\author{H. Sellier, F. Lefloch}
\affiliation{D\'epartement de Recherche Fondamentale sur la Mati\`ere Condens\'ee, CEA Grenoble, 17 rue des Martyrs, 38054 Grenoble Cedex, France.}

\author{M. Faur\'e}
\affiliation{Centre de Physique Mol\'{e}culaire Optique et Hertzienne, Universit\'{e} de Bordeaux 1 and CNRS, 33405 Talence Cedex, France.}

\author{A. Buzdin}
\affiliation{Centre de Physique Mol\'{e}culaire Optique et Hertzienne, Universit\'{e} de Bordeaux 1 and CNRS, 33405 Talence Cedex, France ; Institut Universitaire de France.}

\author{H. Courtois} 
\affiliation{Centre de Recherches sur les Tr\`es Basses Temp\'eratures - C.N.R.S.  and Universit\'e Joseph Fourier, 25 Avenue des Martyrs, 38042 Grenoble, France ; Institut Universitaire de France.}

\date{\today}

\begin{abstract}
We studied the proximity effect between a superconductor (Nb) and a diluted ferromagnetic alloy (CuNi) in a bilayer geometry. We measured the local density of states on top of the ferromagnetic layer, which thickness varies on each sample, with a very low temperature Scanning Tunneling Microscope. The  measured spectra display a very high homogeneity. The analysis of the experimental data shows the need to take into  account an additional scattering mechanism. By including in the Usadel equations the effect of the spin relaxation in the ferromagnetic alloy, we obtain a good description of  the experimental data.
\end{abstract}

                              
\maketitle

\section{Introduction}

Hybrid systems made of a ferromagnetic metal (F) in contact with a superconductor (S) exhibit a rich variety of physical effects. At the interface, the Andreev reflection \cite{Andreev,SaintJames} is responsible for the appearance of the proximity superconductivity in the ferromagnetic metal. In the ferromagnetic metal, the reflection of a spin up (down) electron into a spin up (down) hole correlates the occupancy of the two electron states of opposite spin bands. This process is equivalent to the creation of an electron pair (an Andreev pair). In the framework of a Stoner model for the ferromagnetic metal, spin up and spin down energy bands are shifted by the exchange field $E_{ex}$. This implies that the paired electrons states have a different wave-vector. At the Fermi level, the wave-vector mismatch is $\delta k = k_F E_{ex} / E_F$ where $k_F$ and $E_F$ are the Fermi wave-vector and energy. The Andreev pairs have therefore a non-zero momentum, which induces an oscillation of the pair density as a function of the distance to the interface \cite{Buzdin1,Buzdin2,Demler}. In a diffusive ferromagnetic metal, the oscillation goes together with a decay on the same length scale, i.e. the ferromagnetic coherence length $\xi_F = \sqrt{\hbar D_F/E_{ex}}$ where $D_F$ is the diffusion coefficient in F. The electron energy spreading contributes also to the wave-vector mismatch, but usually in a negligible way, since the exchange field $E_{ex}$ is usually much larger than the thermal energy $k_B T$. 

Depending on the distance to the interface, the superconducting wave-function phase changes from $0$ to  $\pi$. In a S/F/S junction, a $0$-phase or a $\pi$-phase regime is expected, depending on the ferromagnetic metal length as compared to the coherence length $\xi_F$. The crossover between a $0$-phase and a $\pi$-phase regime was observed in Nb/CuNi/Nb junctions \cite{Ryazanov1,Sellier}. There, the thermal energy is not negligible as compared to the exchange field and a sign change of the supercurrent was observed when the temperature was varied. In Nb/PdNi/Nb junctions, the critical current does not change sign with temperature but still exhibits an oscillating behavior as a function of the ferromagnet thickness \cite{Kontos2}.   A $\pi$-regime was also detected in an interference device made of five $\pi$-junctions \cite{Ryazanov2} and in a SQUID made of one 0-junction and one $\pi$-junction \cite{Guichard}.

The $\pi$-regime can also be observed through density of states measurements in a S/F junction. The first observation was performed with solid tunnel junctions deposited on top of Nb/PdNi bilayers \cite{Kontos1}. For a certain range of PdNi thickness, the density of states appeared as reversed, with a maximum at the Fermi level and a minimum at the gap energy. The data were fitted with the quasi-classical theory \cite{Zareyan}. The density of states modulation appeared to be surprisingly small, and shifted to large ferromagnetic alloy thickness. The latter fact was accounted for by the possible existence of a magnetically-dead layer at the interface. The spatial homogeneity of the observed density of states was not investigated because of the intrinsic spatial averaging of the technique. This leaves open questions that motivated us to study similar samples with a very low temperature scanning tunneling microscope (STM). The spatial resolution should  constitute a clear benefit of this new technique, while the energy resolution is comparable.

\section{Experimental study}

\begin{figure} [t]
\includegraphics*[scale=0.5]{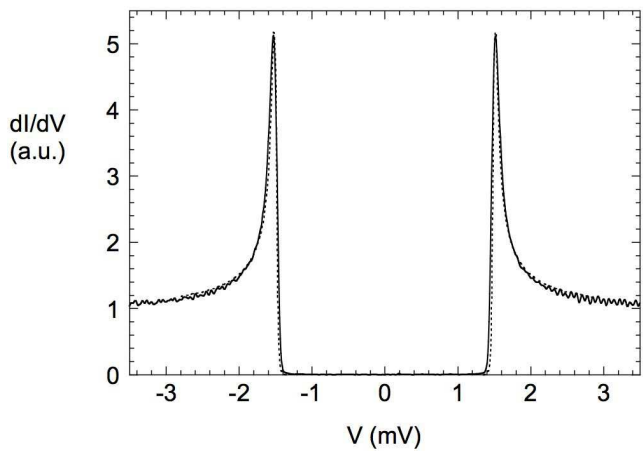}
\caption{\label{Resolution} Spectrum measured on a plain Nb layer (full line) and a fit with a BCS calculation (dashed line). The fit parameters are the energy gap $\Delta=1.5\, meV$ and an effective electronic temperature of 170 mK.}
\end{figure}

Our study relies on six different wafers made with the same schematics. On every silicon wafer, we deposited first a thick layer of Nb and then a thin layer of a CuNi alloy. The CuNi target composition was Cu$_{50}$Ni$_{50}$. The sputtering conditions were identical to that of  Ref. \cite{Sellier}, where Nb/CuNi/Nb junctions critical current was studied.  The Nb layer thickness of 200 nm (at the wafer center) was chosen to be much larger than the Ginzburg-Landau coherence length which is estimated to be 12 nm. The CuNi film thickness varies on each sample (from 2.5 to 20 nm at the wafer center). Because of the characteristics of the chamber, there is a CuNi film thickness gradient from the center to the sides of the wafer. This was taken into account in every indicated thickness in the following. 

CuNi (like PdNi) is a diluted ferromagnetic alloy with a relatively small exchange field. This brings the advantage of a sizable coherence length $\xi_F$ of the order of several nm. On the other hand, the ferromagnetic properties depend strongly on the exact composition of the alloy. The Cu and Ni concentration was measured on one sample by Rutherford BackScattering to be Cu$_{52}$Ni$_{48}$. The Curie temperature of similar films was measured to be about 20 K and the saturation magnetization was extrapolated to $5.10^{4}\, A.m^{-3}$, which corresponds to a magnetic moment of $0.06\, \mu_{B}.atom^{-1}$ (see Ref. \cite{Sellier} for more details). As both the Curie temperature and the magnetic moment depend linearly on the concentration,  an extrapolation \cite{Ahern} for 48$\%$  indicates a value $E_{ex}$ = 12 meV. Taking into account a mean free path in CuNi of about 1 nm, which corresponds to a electron diffusion coefficient $D_F$ of 5 cm$^2$/s, this gives an estimated coherence length $\xi_F$ = 5.0 nm.

An important property of our CuNi films is the absence of hysteresis or remanent magnetization \cite{Sellier}. This property indicates a very weak ferromagnetism close to the superparamagnetic behavior that indeed occurs at 45$\%$ of Ni in bulk alloys \cite{Hicks}. In this regime, the magnetic correlations are short range and the direction of the magnetization varies spatially without domain wall. The characteristic length of these magnetic fluctuations is a few hundred atoms, close to the superconducting coherence length, so that the exchange energy should appear as uniform in terms of induced superconductivity.

\begin{figure} [t]
\includegraphics*[scale=0.5]{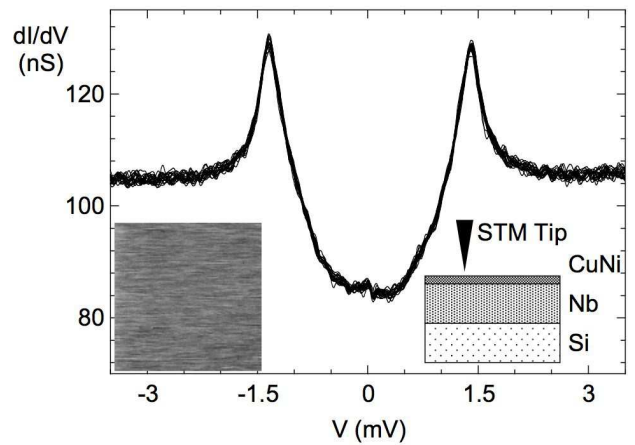}
\caption{\label{Homogeneity} Superposition of $20$ tunneling spectra measured on a sample with a 3.3 nm-thick CuNi film. Spectra were acquired every 11 nm along a single line. The tunnel conductance is of the order of $100\, nS$. Left insert : 400 $\times$ 400 nm$^2$ image of a sample with a CuNi film thickness of 3.3 nm. The black to white scale is 1nm. Right insert : Schematics of the measurement geometry.}
\end{figure}

We used a very low temperature (60 mK) Scanning Tunneling Microscope (STM) developed in our laboratory \cite{STM}. Local spectroscopy measurements were performed at a temperature below 100 mK and with a typical tunnel resistance value of 10 to 25 $M\Omega$. The measured I(V) characteristics (I is the tunnel current, V the voltage applied on the tip) were numerically differenciated so that we obtain the differential conductance dI/dV. In the limit of a zero temperature, this quantity is directly proportional to the local density of states (LDOS) of the sample. In order to test the energy resolution of the STM, we measured tunneling spectra on a plain Nb film, see Fig. \ref{Resolution}. The spectra is fitted with a thermally smeared BCS density of states, without using any inelastic scattering parameter. The fit parameters are the energy gap $\Delta$ = 1.50 meV and an (effective) electron temperature  T = 170 mK. This value represents an improvement as compared to our earlier experiments \cite{STM}. Nevertheless, the difference with the measured sample temperature (below 100 mK) means that some non-thermal noise is present in the STM.

Fig. \ref{Homogeneity} shows the superposition of $20$ spectra all taken on the 3.3 nm of CuNi sample while scanning along a line with a step size of $11\, nm$. All spectra are almost perfectly superimposed. A similar behavior has been observed on every sample. This means that there are no inhomogeneities in the LDOS up to a scale much larger than $\xi_{F}$. This is possible because of the excellent roughness of the samples, of the order 0.5 nm, and the good stability of the CuNi alloy surface.  Fig. \ref{Homogeneity} shows a typical image, of size 400 $\times$ 400 nm$^2$. No clear feature can be observed. 

\begin{figure} [t]
\includegraphics*[scale=0.5]{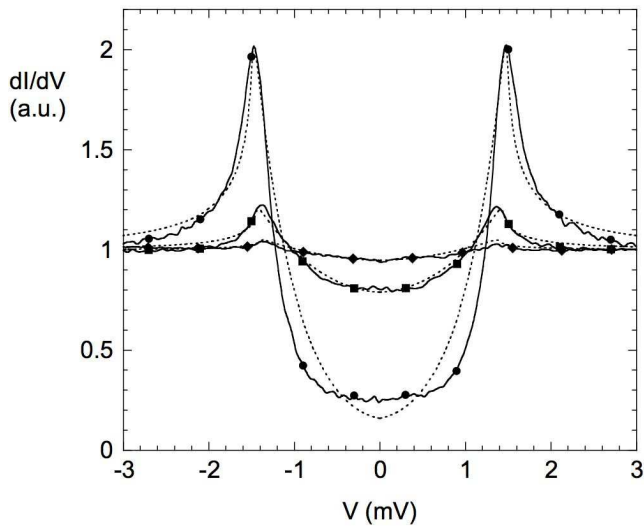}
\caption{\label{Comparison} Tunneling spectra (full lines) measured at the surface of three Nb/CuNi  bilayers with different CuNi thicknesses 2.5 nm (circles), 3.3 nm (squares) and 5.3 nm (diamonds), compared with calculated tunneling spectra (dotted lines). The calculation parameters are $\xi_F$ = 15 nm, T = 170 mK and respectively $L_{s}$ = 3.4 ; 2.6 and 3.0 nm, $\Delta$ = 1.5 ; 1.45 and 1.4 meV.}
\end{figure}

We measured about ten different samples, some of them were taken from the same wafer but at a different place. A selection of data is shown in Fig. \ref{Comparison} for three different thicknesses x of CuNi.  The comparison of these spectra with the Fig. \ref{Resolution} Nb data demonstrates that the smearing of the spectra is not due to a lack of experimental resolution but is a physical effect. For a small thickness of 2.5 nm, there is a sharp rise to a peak in the measured LDOS at $V=\pm \Delta/e$. One can note that $\Delta \simeq$ 1.5 meV, which is very close to the value for bulk Nb (1.6 meV). With a CuNi film thickness of 3.3 nm, the density of states minimum at the Fermi level is less pronounced while the maximum position does not change significantly. This trend appears more strongly in the 5.3 nm sample. As will be discussed below, the behavior of the measured tunneling spectra is not fully monotonous as a function of the CuNi film thickness. For a CuNi thickness larger than 6 nm, where the 0-$\pi$ crossover is expected, the density of states is flat within our experimental accuracy of about 1 $\%$. This limitation comes from the imperfect junction stability that restricts the time averaging possible in a STM experiment.

\begin{figure} [t]
\includegraphics*[scale=0.5]{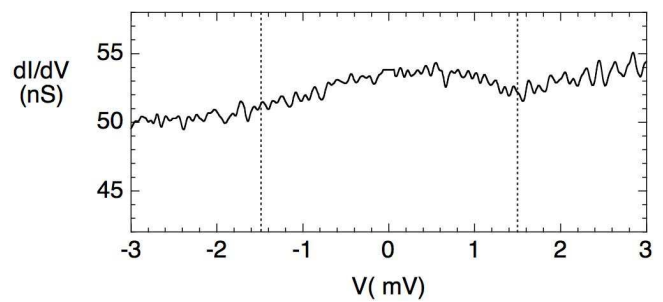}
\caption{\label{PiMax} Normalized differential conductance vs. bias voltage on top of a Nb ($100\, nm$)/ CuNi ($5.1\, nm$) bilayer. Note the scale in $dI/dV$. Experimental noise can be observed. Two dashed lines are drawn at $V = \pm \Delta /e$ for clarity.}
\end{figure}

In one experiment with a CuNi thickness of 5.1 nm, we observed a reversed density of states (see Fig. \ref{PiMax}). This kind of spectra is reminiscent of the $\pi$ regime, which is predicted to appear in a S-F junction at a F metal thickness $x > \pi \xi_F/2$. The background is not as flat as usual. A clear structure is observed at voltages close to $\pm \Delta$ = 1.5 meV (dashed lines), which confirms that this is actually an effect of the proximity with the superconductor. This behavior was reproduced in more than ten successive spectra acquired at different positions, but eventually disappeared as the tip was scanned further. We believe that these spectra did rely on an uncontrolled sample configuration below the tip. For instance the tip may have picked up some CuNi cluster at the top of the sample, so that we performed spectroscopy through a S-F-I-F junction (I for "insulator"). Additional experiments in other samples within the same thickness range did not reproduce the same behavior.

\section{Theory}

Let us now discuss the theoretical description of the superconducting proximity effect in a ferromagnetic metal. We will consider the diffusive regime where the elastic mean free path is supposed to be smaller than every characteristic length scale, including the ferromagnetic metal length. In the diffusive regime, the quasi-classical theory for the inhomogenous superconductors reduces to the Usadel equation. Using the usual parametrization of the normal and anomalous Green functions $G=\cos \Theta $ and $F=\sin \Theta $, the Usadel equation is written for $\omega > 0$ as:
\begin{equation}
-\frac{D_{F}}{2}\frac{\partial ^{2}\Theta }{\partial x^{2}}+\left( \omega+iE_{ex}\right) \sin \Theta +\frac{1}{\tau_{s}}\sin \Theta \cos \Theta =0\text{.}
\label{debut}
\end{equation}
Here we considered a case where the physical quantities depend only on the coordinate perpendicular to the layers. Compared to the usual expression of the Usadel equation, the magnetic scattering in this case is incorporated by replacing $\omega $ by $\omega +G/\tau_{s}$ \cite{Buzdin3}, where $\tau_{s}$ is the magnetic scattering time. Note that we assume the presence of a relatively strong magnetic uniaxial anisotropy in the ferromagnetic layer. This permits us to neglect the magnetic scattering in the plane perpendicular to the easy axis (which mixes up and down spins in the Green functions). Introducing the dimensionless parameters $\widetilde{\omega }=\omega /E_{ex}$, $\widetilde{\tau_{s}}=\tau_{s}E_{ex}$ and $\widetilde{x}=x/\xi_{F}$, Eq. \ref{debut} reads:
\begin{equation}
-\frac{1}{2}\frac{\partial ^{2}\Theta }{\partial \widetilde{x}^{2}}+\left( \widetilde{\omega }+i\right) \sin \Theta +\frac{1}{\widetilde{\tau_{s}}}\sin \Theta \cos \Theta =0\text{.} 
\label{Usadel}
\end{equation}

For an infinite ferromagnet length, the integration of (\ref{Usadel}) using the boundary conditions: $\Theta \left( \widetilde{x}\rightarrow \infty \right) =0$ and $\left({\displaystyle{\frac{\partial \Theta}{\partial \widetilde{x}}}}\right) _{\widetilde{x}\rightarrow \infty }=0$ allows us to find the exact expression : 
\begin{equation}
G=\cos \Theta =1-\frac{8g}{\left( g+1\right) ^{2}-k^{2}\left( g-1\right) ^{2}}\text{,}
\end{equation}
where $k^{2}={\displaystyle{\frac{1}{(\widetilde{\omega }+i)\widetilde{\tau_{s}}+1}}}$ and $g$ is defined as:
\begin{equation}
\label{effect_s}
g=g_{0}\exp \left( -2\sqrt{2}\widetilde{x}\sqrt{\widetilde{\omega }+i+\frac{1}{\widetilde{\tau_{s}}}}\right) 
\end{equation}
with 
\begin{widetext}
\begin{equation}
g_{0}={\displaystyle{\frac{-G_{0}-3-k^{2}\left( G_{0}-1\right) +2\sqrt{\left( G_{0}-1\right) \left( 2+k^{2}\left( G_{0}-1\right) \right) }}{\left( G_{0}-1\right) \left( 1-k^{2}\right) }}}\text{.}
\end{equation}
\end{widetext}
$G_{0}=G\left( \widetilde{x}=0\right) $ is the normal Green function in the S layer, that is
\begin{equation}
G_{0}=\frac{\omega }{\sqrt{\Delta ^{2}+\omega ^{2}}}.
\end{equation}
This means that we neglected the inverse proximity effect and assumed that the pair amplitude at the interface is equal to the one in the bulk. The density of state is deduced from $G$ performing the analytical continuation: 
\begin{equation}
N\left( \epsilon \right) =N(0) \mathop{\rm Re} G\left( x,\omega \rightarrow i\epsilon \right) 
\end{equation}
where $\epsilon =E-E_{F}$ is the energy referenced to the Fermi level. With no spin relaxation, both the oscillation period and the decay length are given by the coherence length $\pi \xi_F$. As can be seen from Eq. \ref{effect_s}, the spin relaxation influences the oscillations of the LDOS in two ways: it diminishes the damping characteristic length of the oscillations and it also increases the period of the oscillations. In the regime $\widetilde{\tau_{s}} \ll 1$, the period is multiplied by the factor $(1+\frac{1}{2 \widetilde{\tau_{s}}})$ which is larger than one. The decay length is divided by the same factor. In the opposite regime, the period is $\pi \sqrt{2} \xi_F^2/L_{s}$ while the decay length is the spin relaxation length $L_{s} = \sqrt{D_S \tau_{s}}$.

For a finite ferromagnet length $d_{F}$, the boundary conditions become $\Theta \left( \widetilde{x}=\widetilde{d_{F}}\right) =\Theta _{d_{F}}$ and $\left( {\displaystyle{\frac{\partial \Theta}{ \partial \widetilde{x}}}}\right) _{\widetilde{x}=\widetilde{d_{F}}}=0$. A first integral of (\ref{Usadel}) gives the following equation:
\begin{widetext}
\begin{equation}
d\widetilde{x}=\frac{d\Theta }{\sqrt{2\left( \cos \Theta _{d_{F}}-\cos
\Theta \right) \left( 2\left( \omega +i\right) +\frac{1}{\tau _{s}}\left(
\cos \Theta _{d_{F}}+\cos \Theta \right) \right) }}\text{,}  \label{premiereintegrale}
\end{equation}
where $\Theta_{d_{F}}$ is the parametrization function at $d_{F}$. In the case of $\Theta_{d_{F}} \ll 1$, the integration of (\ref{premiereintegrale}) can be analytically performed and we obtain:
\begin{equation}
\Theta _{d_{F}}\sim \frac{8}{\sqrt{1-p^{2}}}\exp \left( -\sqrt{2}\sqrt{i+\frac{1}{\widetilde{\tau_{s}}}+\widetilde{\omega }}\widetilde{d_{F}}\right) \sqrt{\frac{\sqrt{1-p^{2}\sin ^{2}\left( \frac{\Theta _{0}}{2}\right) }-\cos\left( \frac{\Theta _{0}}{2}\right) }{\sqrt{1-p^{2}\sin ^{2}\left( \frac{\Theta _{0}}{2}\right) }+\cos \left( \frac{\Theta _{0}}{2}\right) }}\text{,}
\label{thetadf}
\end{equation}
\end{widetext}
where $p^{2}={\displaystyle{\frac{1}{1+ i \widetilde{\tau_{s}}}}}$. The normal Green function at the end of the F layer is deduced from (\ref{thetadf}) and then, the LDOS can be determined with the analytical continuation.

\section{Discussion}

\begin{figure} [t]
\includegraphics*[scale=0.5]{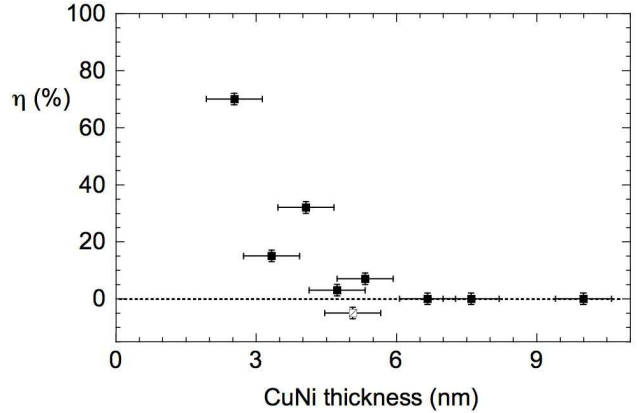}
\caption{\label{Ouverture} Evolution of the change of Fermi level density of states as a function of the samples CuNi film thickness. A value of 100$\%$ for the parameter $\eta$ would correspond to a gap in the density of states. Data are shown with a filled symbol, except for the reversed density of states data (see text) shown with an open symbol.}
\end{figure}

Fig. \ref{Ouverture} shows the LDOS change at the Fermi level $\eta$  as a function of the samples CuNi film thickness. A value of 100 $\%$ for $\eta$ would mean a zero density of states at the Fermi level, while the normal state gives $\eta$ = 0. There is a significant scatter in the data, which cannot be explained by the error bars on the CuNi thickness. The overall behavior is a sharp decrease on a length scale below 3 nm, with no visible $\pi$-regime. This decay length is significantly smaller than the expected coherence length $\xi_F$. This shows that our experimental results do not follow the predictions of the usual theory for F-S junctions.

We added with a different symbol the data point of the $\pi$-regime-like spectra observation.   The reversed density of states observed in one experiment may be understood as a special realization of a $\pi$-regime due to the tunneling through a nanometric ferromagnetic grain weakly coupled to the sample surface. It was actually predicted that a weakly transparent interface in a S-I-F junction can induce a $\pi$-regime in a very thin ferromagnetic layer \cite{Buzdin4}. The physical explanation is that the pair amplitude drop at the interface mimics the effect of a ferromagnetic metal of some thickness. No spin relaxation is expected in the barrier, so that the $\pi$-regime could be of stronger amplitude than in the transparent interface case.

When the F layer is rather small ($2.5\, nm$), $\eta$ is higher than $70\%$. If one extrapolates $\eta$ at the interface ($0\, nm$ of CuNi), one can note that it would reach a value close to $100\%$. This implies that the interface transparency is very good, since a moderate interface transparency would result in a pair amplitude mismatch, and consequently a density of states aperture mismatch. 

We interpret our data by taking into account a large spin relaxation in the CuNi film. This effect was invoked for fitting the critical current measurements results in S/F/S junctions based on the same CuNi layers \cite{Sellier}. In qualitative terms, a spin relaxation length $L_{s}$ shorter than the coherence length $\xi_{F}$ will be the leading term in the decay of the pair amplitude with the distance to the interface. The spin relaxation length estimated in \cite{Sellier} $L_{s}$ = 2.7 nm is at least in qualitative agreement with our experimental findings. 

\begin{figure} [t]
\includegraphics*[scale=0.5]{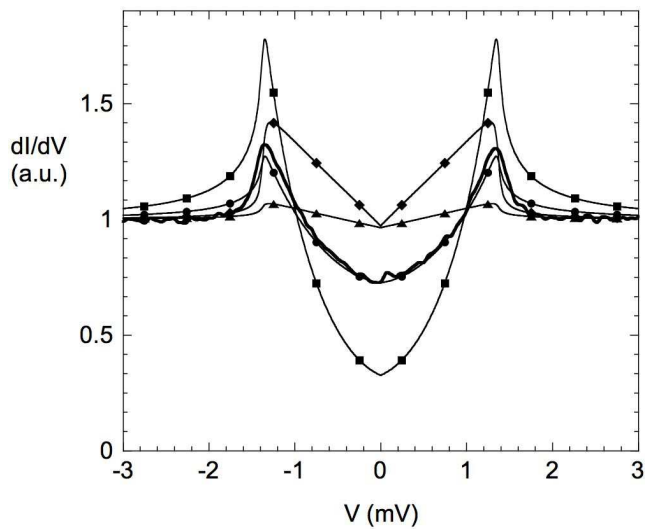}
\caption{\label{Theory} Comparison of calculated density of states spectra (thin lines) with the experimental for the 4.1 nm sample (thick line). The calculation parameters are :  $\Delta$ = 1.37 meV, T = 170 mK, $\xi_F$ ; $L_{s}$) = (15 ; 5) in nm with square symbols ; (15 ; 3.5) with circles ; (4.1 ; 2.5) with triangles and (4.1 ; 4.5) with diamonds.}
\end{figure}

Figure \ref{Theory} shows an experimental spectra of the 4.1 nm sample together with several calculated spectra. The density of states was calculated using Formula \ref{thetadf} and convoluted with a thermal window in order to obtain the predicted tunneling spectra. The energy gap was taken as $\Delta$ = 1.37 meV and the temperature taken equal to 170 mK. The two free parameters were the ferromagnetic metal coherence length $\xi_F$ and the spin relaxation length $L_{s}$. If we take the parameters values that were derived from Ref. \cite{Sellier} critical current measurements ($L_{s}$=2.5 nm and $\xi_F$=4.1 nm), one obtains a calculated spectra (triangle symbols) with an modulation amplitude that is much smaller than in the experimental spectra. Increasing the spin relaxation length ($L_{s}$ = 4.5 nm, cross symbols) enhances the calculated amplitude but the agreement with the experimental data remains very poor.

In contrast, a smaller exchange field gives a good fit to the experimental data. We obtain the best agreement with a spin relaxation length $L_{s}$ = 3.6 nm and a coherence length $\xi_F$ = 15 nm (circle symbols). The latter value corresponds to an exchange field $E_{ex}$ of 1.5 meV, well below the expected value. We interpret this as a signature of the weakened magnetism of the CuNi material in the vicinity of the Nb interface. Since we investigated mostly very thin CuNi films (below than 6 nm), this statement is compatible with the observation of a $\pi$ regime crossover at 17 nm \cite{Sellier}, driven by the larger exchange field of thicker films. A magnetically-dead layer of 1.5 nm at the interface was invoked in PdNi in order to interpret quantitatively density of states measurements \cite{Kontos1}. In the present case of CuNi, we claim that the magnetism of the diluted ferromagnetic alloy is not destroyed but only reduced in the vicinity of the Nb interface. The spin relaxation length $L_{s}$ = 3.6 nm extracted from this fitting procedure is in fair agreement with Ref. \cite{Sellier} and also \cite{CuNi}. Let us note that the calculated spectra depend very strongly on the value of spin relaxation length. For comparison, Fig. \ref{Theory} also shows the calculated spectra with a slightly larger spin relaxation length $L_{s}$ = 5 nm (square symbols). The modulation amplitude is strongly enhanced and the agreement with the data becomes much poorer. 

Fig. \ref{Comparison} shows the comparison between experimental data and the related best fit calculated curves. The coherence length parameter was kept constant with $\xi_F$ = 15 nm, while the spin relaxation length and the energy gap were adjusted in every case. The agreement is very good. The 2.5 nm data is less accurately described, maybe because a lack of validity of the diffusive regime approximation. In this sample, the CuNi thickness is actually of the order of the elastic mean free path. A small variation of about $\pm$ 12 $\%$ in the spin relaxation length is sufficient to account for the scattering of the data. This variation is present even between samples from the same wafer (for instance the 3.3 and 4.1 nm sample). It may be related to small differences in the chemical composition of the CuNi alloy (below the RBS sensitivity of 1$\%$) on a scale larger than the scanning area of the STM. Such small differences can have a dramatic effect on the spin relaxation, especially in the concentration region close to the super-paramagnetic regime.

The energy gap value extracted from the fitting procedure shows a tendency to decrease as the ferromagnetic metal length increases. This small effect may be understood as a small reduction of the energy gap at the F-S interface due to the inverse proximity effect.

\section{Conclusion}

In summary, we investigated experimentally the local density of states in a thin CuNi film in proximity with a Nb film. Our results are consistent with previous studies and bring additional insight into this system. The measured tunneling spectra were very homogenous over the accessible sample area. This may be related to the fact that no magnetic domain structure is expected. The steady evolution of the measured spectra with the CuNi film thickness demonstrates that the interface is very transparent.

The inclusion of the spin relaxation into the Usadel equation modifies the predicted LDOS in a significant way. Comparing our experimental data with theoretical prediction, we find that the spin relaxation has an essential effect on the proximity superconductivity induced in a diluted ferromagnetic alloy like CuNi. This mechanism is responsible for the very small amplitude of the density of states modulation in the $\pi$ regime. The observed data scattering is accounted for by small sample to sample variation of the spin relaxation strength. Moreover, we confirm that, in the vicinity of the Nb interface, the CuNi ferromagnetism is significantically reduced.

\section{Acknowledgements}

Samples were fabricated using the PROMES clean room facilities at the DRFMC (CEA/Grenoble). We would like to thank O. Fruchart for the AFM microscopy, A. Sulpice for SQUID measurements, B. Pannetier, C. Baraduc, V. Falko and M. Houzet for useful discussions and D. Jalabert for RBS measurements. This work was funded by the EU via project NMP2-CT-2003-505587 'SFINx'. We acknowledge support from the ESF "Pi-Shift" project and from the "Action Concert\'ee Nanosciences" program.


\begin{thebibliography}{00}
\bibitem{Andreev} A. F. Andreev, Sov. Phys. J.E.T.P. {\bf 19}, 1228 (1964).
\bibitem{SaintJames} D. Saint-James and P. G. de Gennes, Phys. Lett.  {\bf 4}, 151 (1963).
\bibitem{Buzdin1} A. I. Buzdin, L. N. Bulaevskii, S. V. Panyukov, J.E.T.P. Lett. {\bf 35}, 147 (1982).	
\bibitem{Buzdin2} A. I. Buzdin, M. V. Kuprianov, J.E.T.P. Lett. {\bf 52}, 487 (1990).
\bibitem{Demler} E. A. Demler, G. B. Arnold, M. R. Beasley,  Phys. Rev. B {\bf 55}, 15174 (1997).
\bibitem{Ryazanov1} V. V. Ryazanov, V. A. Oboznov, A. Y. Rusanov, A. V. Veretennikov, A. A. Golubov, J. Aarts, Phys. Rev. Lett. {\bf 86}, 2427 (2001).
\bibitem{Sellier} H. Sellier, C. Baraduc, F. Lefloch, R. Calemczuk, Phys. Rev. B {\bf 68}, 54531 (2003) ; H. Sellier, C. Baraduc, F. Lefloch, R. Calemczuk, Phys. Rev Lett. {\bf 92}, 257005 (2004).
\bibitem{Kontos2} T. Kontos, M. Aprili, J. Lesueur, F. Gen\^et, B. Stephanidis, R. Boursier, Phys. Rev. Lett.  {\bf 89}, 137007 (2002).
\bibitem{Ryazanov2} V. V. Ryazanov, V. A. Oboznov, A. V. Veretennikov, A. Y. Rusanov, Phys. Rev. B {\bf 65}, 20501 (2001).
\bibitem{Guichard} W. Guichard, M. Aprili, O. Bourgeois, T. Kontos, J. Lesueur, and P. Gandit, Phys. Rev. Lett.Ê {\bf 90}, 167001 (2003).
\bibitem{Kontos1} T. Kontos, M. Aprili, J. Lesueur, X. Grison, Phys. Rev. Lett. {\bf 86}, 304 (2001).
\bibitem{Zareyan} M. Zareyan, W. Belzig, Yu. V. Nazarov, Phys. Rev. Lett. {\bf 86}, 308 (2001).
\bibitem{Ahern} S. A. Ahern, M. J. C. Martin, and W. Sucksmith, Proc. Roy. Soc. (London) {\bf 248}, 145 (1958).
\bibitem{Hicks} T. J. Hicks, B. Rainford, J. S. Kouvel, G.G. Low and J. B. Comly, Phys. Rev. Lett. {\bf 22}, 531 (1969).
\bibitem{STM} N. Moussy, H. Courtois, B. Pannetier, Rev. of Sci. Instrum. {\bf 72}, 128 (2001).
\bibitem{Buzdin3} A. Buzdin, Pis'ma Zh. Exp. Teor. {\bf 42}, 283 (1985) [JETP Lett. {\bf 42}, 350 (1985)]
\bibitem{Buzdin4} A. Buzdin, J.E.T.P. Lett. {\bf 78}, 583 (2003).
\bibitem{CuNi} S.-Y. Hsu, P. Holody, R. Loloee, J. M. Rittner, W. P. Pratt and P. A. Schroeder, Phys. Rev. B {\bf 54}, 9027 (1996).
\end{thebibliography}
\end{document}